\documentclass{article}

\PassOptionsToPackage{numbers, compress}{natbib}

\usepackage[preprint]{neurips_2024}

\usepackage[utf8]{inputenc} 
\usepackage[T1]{fontenc}    
\usepackage{hyperref}       
\PassOptionsToPackage{colorlinks}{hyperref}
\PassOptionsToPackage{pdftex}{hyperref}
\usepackage{url}            
\usepackage{booktabs}       
\usepackage{amsfonts}       
\usepackage{nicefrac}       
\usepackage{microtype}      
\usepackage{xcolor}         
\usepackage{graphicx}
\usepackage{xspace}
\usepackage{enumitem}
\usepackage{verbatim}
\usepackage{stmaryrd} 
\usepackage{subcaption} 
\usepackage{relsize}

\definecolor{figurecolor}{RGB}{22,90,220}
\definecolor{citecolor}{RGB}{198,81,19}
\usepackage{caption} 
\usepackage[nameinlink,noabbrev]{cleveref} 
\def\Snospace~{\S{}}

\hypersetup{ colorlinks=true, urlcolor=black, linkcolor=figurecolor, citecolor=citecolor, pdfstartview=FitH
        pdftitle={Cloud Atlas: Efficient Fault Localization for Cloud Systems using Language Models and Causal Insight}
        }

\newcommand{\sys}{Atlas\xspace}

\title{Cloud Atlas: Efficient Fault Localization for Cloud Systems using Language Models and Causal Insight
}

\author{%
  Zhiqiang Xie\thanks{Work partially done during internship at Microsoft Research.} \\
  Stanford University\\
  \texttt{xiezhq@cs.stanford.edu} \\
  \And
  Yujia Zheng \\
  Carnegie Mellon University \\
  \texttt{yujiazh@cmu.edu} \\
  \And
  Lizi Ottens \\
  Stanford University \\
  \texttt{lottens@stanford.edu} \\
  \And
  Kun Zhang \\
  Carnegie Mellon University \\
  \texttt{kunz1@cmu.edu} \\  
  \And
  Christos Kozyrakis \\
  Stanford University \\
  \texttt{christos@cs.stanford.edu} \\  
  \And
  Jonathan Mace \\
  Microsoft Research \\
  \texttt{jonathanmace@microsoft.com} \\
}

\begin{document}

\maketitle

\begin{abstract}
    Runtime failure and performance degradation is commonplace in modern cloud systems.  For cloud providers, automatically determining the root cause of {incidents} is paramount to ensuring high reliability and availability as prompt fault localization can enable faster diagnosis and triage for timely resolution.
    A compelling solution explored in recent work is causal reasoning using causal graphs to capture relationships between varied cloud system performance metrics.
    To be effective, however, systems developers must correctly define the causal graph of their system, which is a time-consuming, brittle, and challenging task that increases in difficulty for large and dynamic systems and requires domain expertise. Alternatively, automated data-driven approaches have limited efficacy for cloud systems due to the inherent rarity of incidents.
In this work, we present \sys, a novel approach to automatically synthesizing causal graphs for cloud systems.  \sys leverages large language models (LLMs) to generate causal graphs using system documentation, telemetry, and deployment feedback. \sys is complementary to data-driven causal discovery techniques, and we further enhance \sys with a data-driven validation step.
We evaluate \sys across a range of fault localization scenarios and demonstrate that \sys is capable of generating causal graphs in a scalable and generalizable manner, with performance that far surpasses that of data-driven algorithms and is commensurate to the ground-truth baseline.
\end{abstract}
\section{Introduction}
\label{sec:intro}
Modern cloud systems are large-scale, complex, and dynamic, combining many inter-operating services. Runtime failure and performance degradation occurs frequently, arising for many reasons including software bugs, hardware failures, incorrect high-level design and unanticipated workload dynamics~\cite{ghosh2022how}. To combat such incidents, engineers continuously monitor the health of cloud systems~\cite{kaldor2017canopy, opentelemetry} using detailed telemetry of data emitted by the systems. An incident is triggered when a top-level health metric, such as end-to-end request latency or API error rate, exceeds an acceptable threshold; on-call engineers must then investigate, localize, and mitigate the incident as quickly as possible. This task requires domain-specific expertise supplemented by the on-call engineer's knowledge of the system to understand the root cause of the issue in need of triage.

Automatically localizing faults and determining the root cause of incidents is a goal of utmost importance for cloud providers. A solution to this problem must overcome two fundamental challenges: first, measurement data is diverse; systems emit an enormous amount of data at different layers, including both standardized measurements such as CPU utilization and application-specific measures; second, symptoms of an incident typically cascade from the root cause to all dependent components recursively, resulting in many changes to correlated metrics which may be obfuscated by fault-tolerance mechanisms that change the workload dynamics of the system.

Recent work in the systems community has identified a compelling solution to both challenges: reasoning on causal graphs~\cite{pearl2009causality, ji2023perfce, wang2019grano, liu2023pyrca}.  Yet despite the effectiveness of causal reasoning, it is critically dependent on the correctness of the causal graphs.  Users are faced with an inherent trade-off between rules-based or data-driven construction of causal graphs: 
while rules-based construction often leads to better causal graphs, composing and updating rules manually is an arduous task; 
on the other hand, automated data-driven approaches rely heavily on the specific parametric form of the data distribution and in real systems these assumptions are often not testable or rarely satisfied. Moreover, anomalous system behavior is rare and some system behaviors may never have been observed before. In both cases, the challenge is exacerbated due to the scale of real systems.

We view the construction of causal graphs as a process of converting system knowledge into a structured representation. While some existing data such as dependency graphs can potentially be processed automatically, most critical semantics about system components and measurements only exist informally, i.e. high-level system descriptions are commonplace in the form of design and architecture documents, and primarily exist for the benefit of human consumption.

In this work we present \sys, a tool that uses large language models (LLMs) to automate the understanding and processing of unstructured semantic information into the causal construction process.
Despite their scale, cloud systems are inherently modular, with complexity arising due to composition, encapsulation, and abstraction.  \sys exploits these characteristics by decomposing a system into its constituent components, interpreting textual component descriptions and identifying pairwise causal relationships between measurements.  \sys contains a novel representation of system components and measurements and employs several LLM prompting strategies to exploit this representation.  After constructing a candidate causal graph, \sys applies a novel data-driven validation step to identify possible mistakes in the causal graph and to propose potentially missed connections through Markov blanket discovery.

We validate \sys using real software system datasets and compare the generated causal graphs with those produced by traditional causal discovery algorithms. We further evaluate \sys on several diverse fault localization case studies.  Our results demonstrate that \sys can produce significantly higher-quality causal graphs than existing data-driven techniques, and when used to localize faults, the graphs produced by \sys have comparable effectiveness to the ground-truth causal graphs.
To the best of our knowledge, this is the first study to employ LLMs in the construction of causal graphs capable of localizing faults for large-scale software systems.
In addition to the system itself, we will also release the simulator code and generated measurement dataset as we believe these resources will benefit the broader community.

\section{Background}
\label{sec:background}

\textbf{Fault Localization.} Most incident investigations today are a manual process overseen or driven by a human operator, with little automation~\cite{kaldor2017canopy, opentelemetry}.   Human-driven investigations implicitly leverage the human's mental model of system behavior to navigate the space of recorded metrics and system components.  From experience, code, and familiarity with system design, humans understand the relationships between different metrics and reason using their system knowledge to determine what to investigate via a chain of cause and effect.

\textbf{Causal Reasoning.}
Despite promising results in controlled environments, little automation is deployed in practice: data-driven approaches are inhibited by high dimensionality, excessive correlations, and high integration overhead~\cite{newcombe2014use}.
As a possible solution, causal reasoning is compelling, because it scopes measurements based on their direct causal influences~\cite{pearl2009causality, ji2023perfce, wang2019grano, liu2023pyrca} and because it incorporates missing cause-and-effect domain knowledge that human-driven investigations utilize.  However, capturing and maintaining this knowledge in an up-to-date causal graph has high overhead; consequently causal reasoning is often limited to constrained scenarios where the semantics of a small set of measurements are predefined and combined with predefined rules.

\textbf{Rule-Based Construction.}  In this approach, a domain expert (e.g. a system developer or operator) provides the causal graph by enumerating the causal relationships between variables using their domain knowledge.  In computer systems, the breadth of system diversity and measurements is too large to expect humans to provide the full, correct graph manually.  Though some work~\cite{causil, li2022causal} has explored easing this burden for constrained scenarios, it remains a significant obstacle in practice.

\textbf{Data-Driven Causal Discovery.}  Causal discovery algorithms are a data-driven approach to extracting causal relationships from observed measurements, potentially overcoming the need for time-consuming input from domain experts.  Causal discovery provably cannot produce a single correct result, but can produce viable candidates for evaluation by a human.  Causal discovery is a poor fit for identifying a cloud system's causal structure, because incidents are edge-case system behaviors that are not exercised in the steady-state; thus steady-state telemetry is insufficient for identifying the full diversity of system behaviors.
Acquiring such a dataset is extraordinarily challenging, requiring both the ability to anticipate edge-case behavior that may never have been seen before, as well as the ability to perform interventional testing by running the system at scale and injecting anomalies.  

\section{Observations}
\label{sec:observation}

\sys automates \emph{rules-based construction} of causal graphs for computer system telemetry.
This section describes the key intuitions behind \sys; in \autoref{sec:design}, we present \sys's end-to-end design.

\textbf{Domain Knowledge from System Documentation.}
Causal relationships between system telemetry derive from a high-level model of system components and interactions.  In practice, it is exceedingly rare for such a model to be explicitly formalized.  Instead, developers predominantly communicate the high-level behavior of a system through design and architecture documents, describing in text the system components, topology, and interactions.  Examples can be found in many open-source software projects~\cite{hdfs_user_guide,redis_cluster_architecture}.   System descriptions use a high level of abstraction and make heavy reference to well-known systems concepts such as replication, load balancing, containers, processes, etc.  
Telemetry can be similarly related to system concepts: measurements use descriptive names and are typically accompanied by descriptions of what is being measured.
\sys leverages this observation by utilizing LLMs to interpret text-based system descriptions to extract causal relationships.  LLMs have knowledge of common systems concepts and are able to interpret design documents.  

\textbf{Causal Relationships from System Interactions.}
Although systems can expose a large amount of telemetry, the relationships between measurements are not arbitrary.  We observe that direct causal relationships between telemetry only arise when there is an interaction between corresponding components.  For example, the throughput of service $A$ running on server $X$ can only have a direct causal influence on CPU$_{X}$ utilization because $A$ is directly interacting with the CPU by running instructions on it.  By contrast, there can be no \emph{direct} influence on server $Y$'s CPU$_{Y}$ utilization; that dependency can only be established indirectly through other intermediate components, such as some service $B$ running on $Y$.  Moreover, software systems are designed in a modular, hierarchical manner to mitigate complexity; it is universally discouraged to tightly couple software components.
Therefore, we only need to consider causal relationships between measurements from components that directly interact with one another.  
\sys leverages this observation from 
system documentation, which typically describes directly which components interact with others, as well as
distributed trace data~\cite{opentelemetry} to directly observe runtime interactions.

\textbf{Scalability from Decomposition.}
Thanks to the locality of interaction and causal relationships, \sys does not need to present an LLM with wide-ranging information about the system to identify causal relationships. Instead, it can decompose the task into smaller, manageable pieces, inspecting locally connected components and their relationships independently.  \sys is thus scalable by design; as the size of a system scales up, it increases the number of LLM interactions, but not the complexity of inputs or outputs to and from the LLM for each iteration.

\textbf{Reconciling with the Ground Truth.}
\label{sub:reconciling}
Systems do not measure everything that can truly be measured.  For example, RPC servers in general might measure throughput, queueing time, latency, and request success; yet a specific RPC server implementation might only measure a subset of those.  This poses a potential challenge to correctly identifying causal relationships: two measurements that have an indirect causal relationship in the idealized version of the system might now have a direct causal relationship in a particular system implementation if the intermediate variables aren't observed.  \sys is robust to this by construction: it does not initially work with the actual measurements available in the system, but instead leverages general knowledge of systems concepts to construct the ideal causal graph which contains a superset of the actual measurements available.  \sys then reconciles the ideal causal graph with the actual measurements that are available.

\textbf{Complementary Data-driven Validation}
Unlike data-driven causal discovery, LLMs lack useful indicators like confidence levels, making it difficult for users to identify, localize, and correct any potential errors in the generated causal graph.
Our intuition to address this is that \sys is compatible with data-driven techniques: traditional statistical methods can optionally be leveraged to validate the output of \sys, if desired.  These methods can use measurement data to negate proposed connections that are contradicted by data, and suggest potentially missed connections with confidence scores.  \sys leverages this as an optional additional validation step after generating a causal graph.

\section{Design}
\label{sec:design}

\begin{figure}[t]%
\centering%
\hspace{17mm}%
\includegraphics[width=0.9\textwidth]{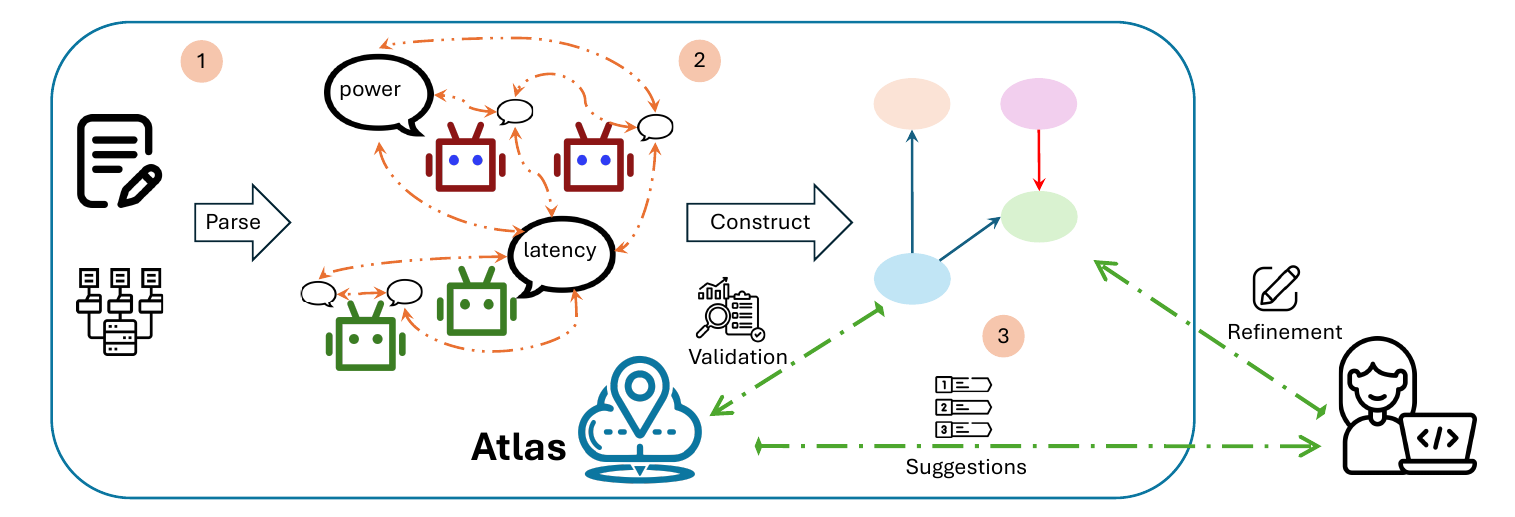}%
\caption{Overview of \sys: (1) Instantiating agents for each system component by parsing system documentation and telemetry; (2) Enumerating causal relationships between metrics; (3) Constructing the causal graph and refining with human-in-the-loop feedback.}%
\label{fig:atlas}%
\end{figure}

\textbf{Overview.}\hspace{1mm} \autoref{fig:atlas} depicts the end-to-end pipeline of \sys.  \sys processes descriptions of system components, metrics, and caller-callee relationships, as well as an optional corpus of common component and metric descriptions. Using an agent-based approach, \sys represents each system component as a separate LLM agent with relevant descriptions. \sys iteratively enumerates all possible metrics for each component and prompts the LLM accordingly to identify causal relationships and their directionality between interacting agents. It then combines these outputs into a single graph, collapsing indirect causal relationships through unmeasured nodes. Finally, \sys presents the candidate causal graphs to users and provides an interface for interactively refining the candidate graph through review and feedback on a small set of proposed causal relationships.

\subsection{Inputs}

The inputs to \sys are organized as follows: 

\begin{itemize}[noitemsep,topsep=0pt,parsep=0pt,partopsep=0pt,leftmargin=20pt]
\item A collection of named system components (e.g. services, processes, etc.) and corresponding textual descriptions.
\item A collection of named measurements and corresponding text descriptions where each measurement is associated with a system component.
\item A list of caller-callee relationships between components.
\item A list of computational resources available to components with corresponding text descriptions.
\end{itemize}
Descriptions can be concise or verbose; for example, the description of a gateway server may simply be, \emph{``Website gateway receives the client request.''}
See \autoref{appendix:input} for detailed examples.
Note most of these inputs can be automatically extracted from request traces, system deployment configurations files, and developer documentation.

\subsection{Instantiating Agents}

\sys conceptualizes each component in the system as an agent.  Agents are categorized as one of the following classes: \emph{request}, \emph{service}, or \emph{resource}.  Each agent is assigned relevant information about its role: component descriptions, metric descriptions, and resource descriptions.  Each agent is also assigned the information about any components it directly interacts with; only the following kinds of interaction are considered: (a) a request invokes a service; (b) a request utilizes a resource; (c) a service utilizes a resource; (d) a service invokes another service. Each agent corresponds to a bounding box in \autoref{fig:two_graphs}.

\subsection{Metrics Enumeration: Creating Nodes for the Causal Graph}
As motivated in \autoref{sub:reconciling}, in addition to having all available metrics of system components as nodes within the causal graph (blue nodes in \autoref{fig:raw}), we also enumerate typical measurements associated with those components and convert them to be nodes as indicated in orange in \autoref{fig:raw}.

\subsection{Causal Relationship Examination: Connecting the Nodes}

\sys iteratively evaluates whether a causal relationship exists between pairs of measurements.  \sys only considers measurements that either (a) exist within the same component; or (b) exist on components that directly interact with one another.  \sys prompts the LLM from the perspective of one agent: it furnishes the LLM with the information known by that agent about its own role and the pair of measurements, asks whether there exists a causal relationship between the two measurements, and if so, asks what the direction of the causal influence is.  \sys uses common techniques to construct the prompt: chain of thoughts~\cite{cot} and alternating order of options with repeating queries for consistent output~\cite{zheng2024judging}. We provide concrete prompt examples in~\autoref{appendix:input}.

\sys employs a semantic cache to store the full semantics of each query and its result.  This optimization greatly reduces the number of times \sys prompts the LLM because the same set of components can be instantiated in multiple places in large-scale systems, with the same semantics and thus same causal relationships between measurements.  By utilizing cached responses, the execution time and token cost of \sys is greatly reduced.

\subsection{Graph Construction}
\label{sub:contraction}

\sys combines the individual causal relationships discovered during the previous step into a causal graph.  The resulting causal graph is over-complete: it includes nodes for all measurements recorded by the system, and can also include nodes for common measurements that \emph{could} be recorded by the system but are not (i.e., unobserved measurements).  We call this graph the \emph{confounder} graph because it exposes potential unobserved confounders; we do not discard the confounder graph because knowledge of potential confounders may be useful for users in some future causal reasoning task or in revisiting the choice of measurements made by the system.

\begin{figure}[t]%
\centering%
\begin{subfigure}[t]{0.5\textwidth}%
\includegraphics[width=\textwidth]{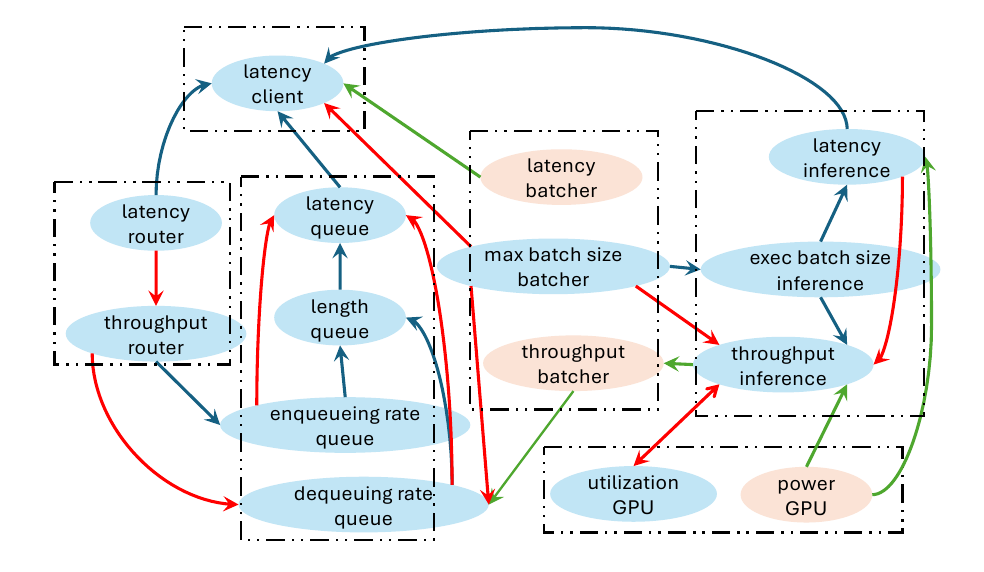}%
\caption{Confounder graph where unobserved nodes are indicated in brown, incorrect causal relationships are indicated with red edges, and reversed directionality is indicated with double arrows.}%
\label{fig:raw}%
\end{subfigure}%
\hspace{0.05\textwidth}%
\begin{subfigure}[t]{0.45\textwidth}%
\includegraphics[width=\textwidth]{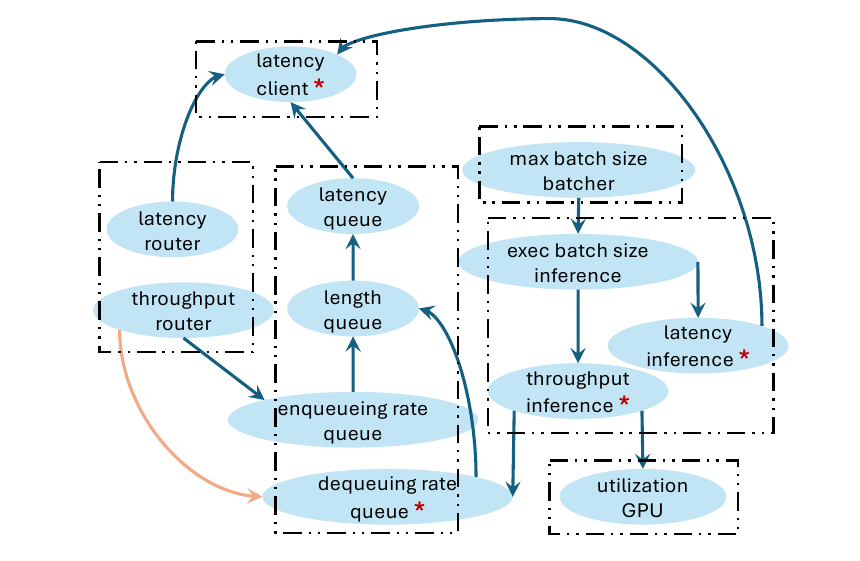}%
\caption{Causal graph after post-processing where the red asterisks indicate that collapsed nodes are recorded. Incorrect edges remaining after refinement are indicated with yellow edges.}%
\label{fig:post}
\end{subfigure}%
\caption{Causal graph of a model serving task (some nodes and edges omitted for readability).}%
\vspace{-1em}%
\label{fig:two_graphs}%
\end{figure}

To produce the final causal graph, \sys duplicates the confounder graph and deletes all unobserved nodes.  To delete an unobserved node, \sys removes the node while preserving any transitive causal influence that may exist, for example, if $A$ and $B$ are observed, but $U$ is unobserved, then $A \shortrightarrow U \shortrightarrow B$ collapses to $A \shortrightarrow B$. 
The locations of these collapsed unobserved nodes are recorded, which will be useful for fault localization as explained in \autoref{sub:case}.
By removing unobserved nodes, causal relationships that were previously indirect may now be direct.  \autoref{fig:two_graphs} illustrates this process.

\subsection{Data-driven Human-in-the-loop Pareto Refinement}
\label{sub:validation}
\sys lacks guarantees that the resulting causal graph is correct, so it employs an optional human-in-the-loop refinement step to identify and correct four kinds of errors that may exist in the graph: cycles, false positives, false negatives, and reversed directionality.  As motivated in \autoref{sec:observation}, we employ data-driven approaches to filter out a small set of candidates for feedback and refinement.
Specifically, we:
1) Apply Markov blanket discovery to examine all proposed causal relationships. We sample five suspicious candidates, prioritizing edges whose removal would change connectivity, and ask the user to correct any erroneous connections. This process is iterative until all five candidates are correct.
2) Apply the Additive Noise Model to examine the direction of established edges, following the same procedure.
If there still exists a cycle in the directed acyclic graph, \sys{} will suggest a minimal set of edges to cut to resolve the issue.
3) Apply Markov blanket discovery again to suggest potentially missed connections among edges that will create new connectivity.
In this optional validation stage, the user only needs to manually examine a small set of causal relationships to obtain a significant Pareto refinement on the causal graph as shown in \autoref{tab:f1}. Shared in
\autoref{fig:post} is a sample causal graph after this refinement procedure.

\section{Evaluation}
\label{sec:eval}

\newcommand{\atlas}{\textsc{\textls[-50]{Atlas}}\xspace}
\newcommand{\atlasfull}{\textsc{\textls[-50]{Atlas\texttt{+}}\textls[-150]{V}}\xspace}
\newcommand{\naivellm}{\textsc{\textls[-50]{Na\"{i}ve}\textls[-100]{LLM}}\xspace}

Our evaluation aims to answer the following key questions:
\begin{itemize}[noitemsep,topsep=0pt,parsep=0pt,partopsep=0pt,leftmargin=20pt]
    \item Does \sys generate superior causal graphs compared to traditional data-driven causal discovery algorithms? (\autoref{sec:eval:comparison})
    \item How effective are the causal graphs generated by \sys for fault localization use cases? (\autoref{sec:eval:localization})
\end{itemize}

In addition to addressing the questions above, our evaluation concludes with three \sys  case studies (\autoref{sec:eval:casestudies}).

\textbf{\sys Variants.} To better understand the contribution of Atlas' design, our evaluation compares three different variants of Atlas: \textbf{\naivellm} uses a one-shot LLM prompt to construct the graph. It provides all of the text descriptions used by Atlas in a single input, and prompts the LLM to generate all edges of the causal graph. \textbf{\atlasfull} utilizes the full \sys pipeline including the data-driven validation step described in~\autoref{sub:validation}. \textbf{\atlas} utilizes the full pipeline, but omits the final data-driven validation step.  All \sys variants use GPT-4-turbo as the LLM.  Due to non-deterministic LLM outputs, we repeat all \sys experiments 5 times and report averaged results.

\textbf{Data-driven Causal Discovery.} We compare the causal graphs generated by \sys to those produced by the following data-driven causal discovery algorithms: GES~\cite{chickering2002optimal}, GRaSP~\cite{lam2022greedy}, and PC~\cite{spirtes2000causation}. 
All experiments are also repeated 5 times and averaged results are reported.

\textbf{Existing Scenarios.}  We make use of several pre-existing datasets from prior work that provide observational data and ground-truth causal graphs for systems scenarios: the Middleware Oriented Message activity (MoM) and Antivirus Activity (AA) scenarios from prior IT system monitoring work~\cite{ait2023case} and the Microservices latency scenario from the DoWhy causal reasoning library~\cite{dowhy_rca}.  We supplement the observational data with brief textual descriptions of the system components.  

\begin{figure}[t]%
\centering%
\vspace{-1em}
\includegraphics[width=0.7\textwidth]{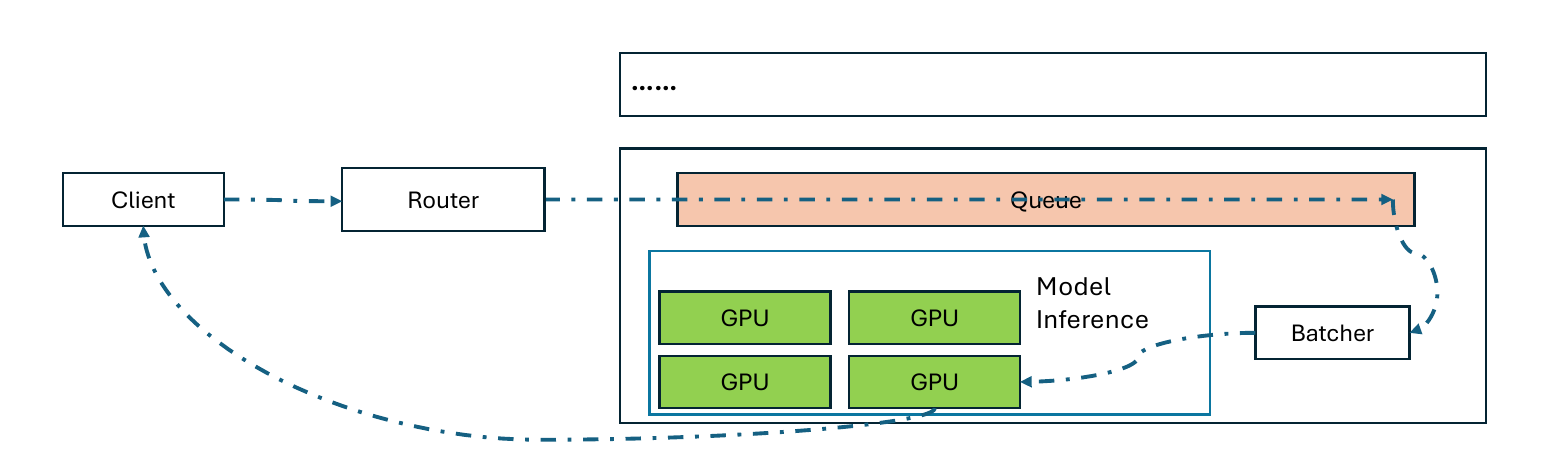}%
\caption{Simulated Model Serving Diagram.}%
\vspace{-1em}
\label{fig:servingsimulator}%
\end{figure}

\textbf{New Scenarios.}  To evaluate \sys at larger scale, we have developed a discrete event simulator that can simulate the dynamics of cloud environments including dynamic service topologies, execution flow, resource contention, workloads, and faults.  Within this simulator we implement a model serving service and workload, following the design of DLIS~\cite{soifer2019deep}.  \autoref{fig:servingsimulator} depicts the high-level execution flow: a client's request reaches a router and is load-balanced to a worker server; it enters a queue, and when a GPU becomes available for inference, the request is dequeued and executed.  Different requests can be batched together for GPU execution to achieve higher throughput.  The system can be configured with a variable number of machines, each with a variable number of GPUs. We simulate workloads with exponentially distributed request interrarival times and permute the following dimensions to generate new datasets:
\begin{itemize}[noitemsep,topsep=0pt,parsep=0pt,partopsep=0pt,leftmargin=20pt]
\item ModelServingS configures one worker machine with one GPU; ModelServingM configures one worker machine with four GPUs; and ModelServingL configures two worker machines each with four GPUs. 
\item For fault localization scenarios we simulate both a normal operating state and several failure scenarios that are commonly observed in model serving systems~\cite{258914}, including workload spikes, network slowdown, batch size misconfiguration, and GPU throttling.  We generate 7, 13, and 24 different failure datasets for ModelServing -S, -M, and -L respectively.
\end{itemize}

Each scenario varies in the size of the ground-truth causal graph and the number of observational samples. \autoref{tab:f1} summarizes these characteristics.

\subsection{Comparison on Causal Graph Construction}
\label{sec:eval:comparison}

In this experiment, we apply all causal graph construction techniques to all datasets.  We measure the effectiveness of each technique by calculating the F1 score of the edges of the constructed graphs with respect to a ground truth baseline.  We present the results in \autoref{tab:f1}.

In general, while different causal discovery algorithms (GES, GRaSP, PC) exhibit unique strengths depending on the data distribution, they generally perform poorly in system troubleshooting scenarios as indicated by the results in~\autoref{tab:f1}. This underperformance is primarily due to a fundamental bias in system measurement data: most data is collected during normal operating states as abnormal states are rare, diverse, and often fail to be recorded. Additionally, these algorithms tend to perform worse as the number of nodes increases.

By contrast, \atlas attains the highest F1 scores across all datasets.  On small datasets, \atlas either generates the same causal graph as the ground-truth, or has additional superfluous edges.  \atlas maintains strong performance as the graph size increases.  \atlasfull demonstrates further improvements over the baseline \atlas: the additional data validation and Pareto refinement steps further improve the quality of the causal graph.  The average number of proposed and accepted graph modifications of \atlasfull is included in \autoref{tab:f1}.

\begin{table}[h!]%
\small%
\centering%
\vspace{-1em}%
\caption{F1 scores of causal graphs compared to ground truth.  We average \sys results over 5 repetitions and report the number of changes proposed and accepted by \atlasfull's validation step.
}\label{tab:causal_discovery_comparison}%
\label{tab:f1}%
\setlength\tabcolsep{5pt}
\begin{tabular}{@{}|l|l|l|l|l|l|l|@{}}
\hline
Method & MoM & AA & \textls[-30]{Microservice} & \textls[-30]{ModelServingS} & \textls[-30]{ModelServingM} & \textls[-30]{ModelServingL} \\
\hline
\hline
\emph{\# Nodes} & 7 & 13 & 11 & 15 & 30 & 56 \\
\emph{\# Edges} & 10 & 16 & 13 & 15 & 36 & 70 \\
\emph{\# Samples} & 654 & 2643 & 10001 & 7087 & 6966 & 13601 \\
\hline
\hline
GES & 0.14 & 0.19 & 0.28 & 0.22 & 0.30 & \emph{timeout} \\
GRaSP & 0.17 & 0.26 & 0.18 & 0.28 & 0.24 & 0.14  \\
PC & 0.29 & 0.20 & 0.37 & 0.28 & 0.29 & 0.12  \\
\hline
\hline
\naivellm & 0.65 & 0.44 & 1.0 & 0.06 & 0.0 & 0.03\\ 
\atlas & 1.0 & 0.86 & 1.0 & 0.65 & 0.68 & 0.69\\
\atlasfull & \textbf{1.0} & \textbf{0.86} & \textbf{1.0} & \textbf{0.89} & \textbf{0.79} & \textbf{0.73}\\  
\hline
\hline
\relsize{-1}{\hspace{2mm}Accepted/Proposed} & \relsize{-1}{--} & \relsize{-1}{0/5} & \relsize{-1}{--} & \relsize{-1}{11.8/14.8} & \relsize{-1}{11.8/16.5} & \relsize{-1}{4.4/7.6}\\        
\hline
\end{tabular}%
\end{table}

\subsection{End-to-end Fault Localization on Model Serving}
\label{sec:eval:localization}

We evaluate the impact of \sys for fault localization use cases, and demonstrate that \sys' superior causal graphs result in significantly improved fault localization performance.  We focus on the three ModelServing scenarios for which we have separate normal-state and failure-state observational datasets.

We utilize an off-the-shelf fault localization algorithm ({\textls[-20]{\relsize{-0.5}{\texttt{distribution\_change}}}}) from the DoWhy causal reasoning library~\cite{dowhy_gcm}.  This algorithm takes as input a causal graph, a dataset of normal-state observations, and a dataset of failure-state observations.  The algorithm calculates and outputs an attribution score for each dataset feature, representing the likelihood that the feature is the root cause.

We apply the algorithm separately using the causal graphs generated by \atlas and \atlasfull, and to the ground-truth causal graph.  To serve as comparison, we also apply the algorithm to the highest scoring causal graph generated by GES, GRaSP, or PC (as reported in~\autoref{sec:eval:comparison}).  We consider a failure to be correctly identified if the true root cause appears in the output and evaluate the top-1 and top-3 ranked features.  We repeat this process for all failure datasets.

\autoref{tab:fault_localization_comparison} reports results averaged across the 7, 13, and 24 failure scenarios for ModelServing -S, -M, and -L respectively.  The results illustrate that the \atlas and \atlasfull graphs are sufficient for identifying root causes across most failure scenarios.  Moreover, they do not perform significantly worse than the ground-truth causal graph.  Upon inspection, we found that \sys graphs were succeeding and failing for the same scenarios as the ground-truth causal graph.  By contrast to \sys, the graphs generated by causal discovery algorithms perform poorly and do not reliably localize faults.  

These results indicate that F1 scores are not a sufficient indicator that a causal graph is effective for fault localization.  We highlight two reasons for this.  First, counterfactual reasoning is only conducted on the subgraph of causal predecessors of the symptom node, whereas the F1 score measures the relative correctness of the entire graph.  Second, a single incorrect edge or inverted directionality can have a minor effect on the F1 score, but a significant effect on the efficacy of causal reasoning.  When examining the causal graphs produced by \sys, we found that most incorrect edges do not change the connectivity of the causal graph and do not break the chain of causality from symptom to root cause, enabling comparable performance to the ground truth.  We examine this effect in more detail in ~\autoref{sub:case}.

\begin{table}[h!]%
\centering%
\small%
\vspace{-1em}
\caption{Percentage of faults successfully localized using different causal graphs.
}\label{tab:fault_localization_comparison}%
\begin{tabular}{|l|l|l|l|l|l|l|}
\hline
& \multicolumn{2}{|l|}{ModelServingS} & \multicolumn{2}{|l|}{ModelServingM} & \multicolumn{2}{|l|}{ModelServingL}\\
\hline Causal Graph & Top 1 & Top 3 & Top 1 & Top 3 & Top 1 &Top 3\\
\hline
Best Causal Discovery Algorithm & 0\% & 14\% & 0\% & 8\% & 25\% & 29\% \\
\atlas & 60\% & 97\% & 90\% & 92\% & 96\% & 96\%\\
\atlasfull & 60\% & 100\% & 90\% & 92\% & 96\% & 96\%\\
Ground Truth & 57\% & 100\% & 92\% & 92\% & 96\% & 96\%\\
\hline
\end{tabular}%
\end{table}
\vspace{-1em}

\subsection{Case Studies}
\label{sub:case}
\label{sec:eval:casestudies}

\begin{figure}[t]%
\centering%
\begin{subfigure}[t]{0.33\textwidth}%
\centering%
\includegraphics[width=1\textwidth]{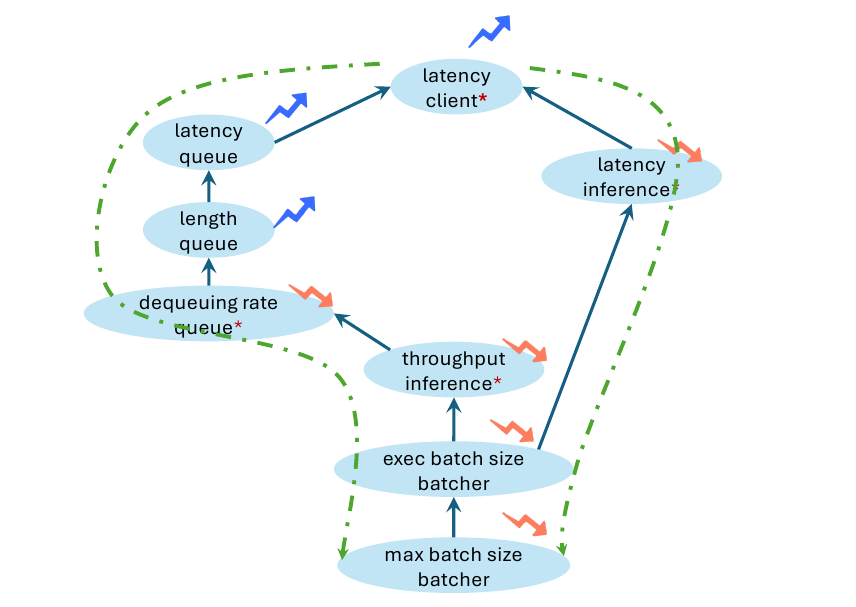}%
\caption{Chain of causal effect for \\
a wrong batch size setting}%
\label{fig:chain}%
\end{subfigure}%
\begin{subfigure}[t]{0.33\textwidth}
\centering
\includegraphics[width=1\textwidth]{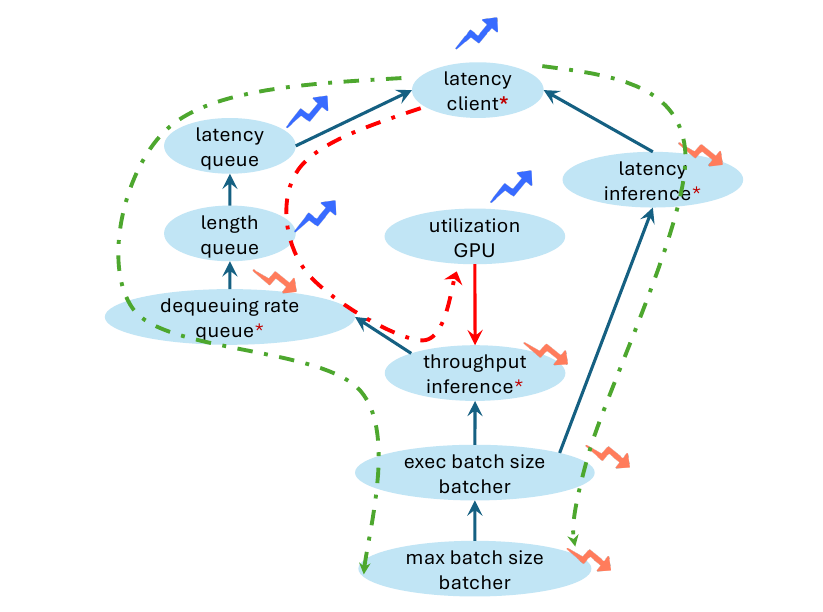}
\caption{Same setting with a \\
reversed edge (in red)}
\label{fig:false_edge}
\end{subfigure}
\begin{subfigure}[t]{0.33\textwidth}%
\centering%
\includegraphics[width=0.75\textwidth]{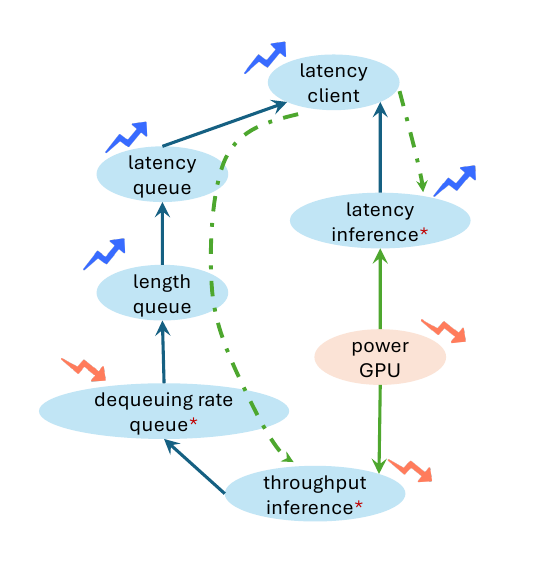}
\caption{Fault localization for the case of missing metric}%
\label{fig:missing_metric}
\end{subfigure}%
\caption{Case study on localizing a fault in ModelServing}%
\label{fig:case_study}%
\vspace{-5mm}%
\end{figure}

We examine the causal graph in the ModelServing scenario to better understand the behavior of fault localization and its limitations with respect to causal graphs.

\textbf{Localize the Real Fault Beyond Discrete Signals.} 
A key strength of causal reasoning compared to other fault localization techniques is its robustness to diverse telemetry signals, including discrete, categorical, and continuous data.  By comparison, many techniques discard useful information by reducing continuous signals to discrete indicators.  For example, recent large-scale training works characterize execution time as ``normal'' and ``too long''~\cite{jiang2024megascale}; in general, techniques based on e.g. decision trees are often reduced to a threshold comparison such as ``latency is above X''.

To illustrate, consider a scenario where an incorrect max batch size is set for the batcher. As illustrated in \autoref{fig:chain}, a smaller batch size reduces effective throughput, causing more requests to queue and thereby increasing queuing latency. Although the reduced batch size decreases model inference latency, overall user latency is longer.
Fault localization techniques might incorrectly report abnormally high queuing latency as the root cause, leaving the developer to diagnose the issue. With counterfactual reasoning, the fault can be traced to its root cause by the chain of causal relations.

\textbf{How a Wrong Causal Relationship Degrades Fault Localization Performance.}
In the same setting, \autoref{fig:false_edge} depicts a false case generated by \sys{}, introducing an inverted causal relationship between GPU utilization and inference service throughput compared to \autoref{fig:post}. This incorrect direction creates a new causal effect chain, inhibiting counterfactual reasoning and lowering the attribution score of batch size. As a result, the max batch size is relegated from the top-3 to top-5 likely root cause.  \sys' validation process described in \autoref{sub:validation} helps identify and correct this erroneous direction, thereby enhancing the accuracy of fault localization.

\textbf{Fault Localization When Metrics are Missing.}
In practice, a system might not report all possible measurements.  \sys is robust to this by generating a confounder graph alongside its output causal graph.  In the Model Serving scenario, GPU power is an example of an unobserved measurement.  Cloud systems often dynamically scale the power draw of GPUs, but we often lack direct measurements of power.  In this case, the graph produced by \sys enables causal reasoning to attribute a fault to latency or inference service throughput as illustrated in \autoref{fig:missing_metric}. From the confounder graph, GPU power will also be reported to the developer as a possible culprit in this case.

\section{Related Work}
\textbf{LLMs for System Troubleshooting.}
RCACopilot~\cite{chen2023automatic} proposed a pipeline that leverages LLMs as powerful text processing tools to summarize historical incidents, match them with newly occurred incidents, and synthesize possible root causes and mitigations based on historical data. This exemplifies a line of research~\cite{roy2024exploring} utilizing LLMs as information processing and reasoning engines. However, the effectiveness of these approaches strongly depends on the historical data, as identical symptoms can arise from entirely different causes, and similar symptoms may have never occurred before.
\sys employs a two-stage approach to overcome this limitation: LLMs are used to process domain knowledge and assist in causal graph construction, while the reasoning is powered by a causal reasoning engine that utilizes abundant measurement data.

\textbf{Causal Discovery.} Causal discovery aims to uncover causal relations from observational data without any interventional experiments. Classical algorithms include constraint-based methods (e.g., PC \citep{spirtes2000causation}), score-based methods (e.g., GES \citep{chickering2013}), and many others (e.g., GRaSP \citep{lam2022greedy}). Recently, large language models (LLMs) have demonstrated impressive empirical performance in addressing some causality-related tasks \citep{kiciman2023causal, zhang2023understanding, naik2023applying, long2023causal, takayama2024integrating, jiralerspong2024efficient, liu2024discovery}. Although LLM-based methods lack comprehensive theoretical guarantees, their notable successes in diverse scenarios highlight their potential applicability in complex real-world tasks, where assumptions for identifiability theory often do not hold perfectly in practice.

\textbf{Causal Reasoning for System Troubleshooting.} Several research works leverage the causal structure of service dependency graphs, where nodes are common service measurements such as latency~\cite{gan2021sage, lin2018microscope, meng2020localizing, wang2019grano, liu2023pyrca}.  This causal structure only relates a small number of externally-visible measurements, however it can serve as a useful starting point for constructing causal graphs, e.g. by including more metrics and their causal relationships~\cite{causil, li2022causal}.  For more diverse measurements such as hardware utilization, service rate, or application-defined metrics, no universal rules or mapping exists.  Causal discovery algorithms have been utilized for microservice metrics~\cite{ikram2022root}, and face the challenges outlined in this paper.  Chaos engineering is a compelling approach to uncover the diversity of system execution behaviors in small scale~\cite{ji2023perfce}; however scaling such an approach to a full cloud system represents a significant challenge.

\section{Limitations and Future Work}
\label{sec:future}

Constructing causal graphs is a key step for effective fault localization.  However, end-to-end there remain several obstacles.  First, causal reasoning algorithms scale poorly to large graphs; we have identified several optimization opportunities based on graph decomposition that we are leveraging in our ongoing work.  Second, causal graphs require special handling for the temporal feedback cycles that exist in systems, which cannot be represented as a DAG.  Third, the causal graph is insufficient for representing all semantics of a system, such as semantic equivalence between components that are represented by different subgraphs.  Lastly, domain knowledge can provide hints as to the correct causal mechanism to assign to each node; this is not currently captured.  Our future work on \sys will be guided by user feedback from open-sourcing the project.

\section{Acknowledgement}
This research was partly supported by the Stanford Platform Lab and its affiliates, and by ACE, one of the seven centers in JUMP 2.0, a Semiconductor Research Corporation (SRC) program sponsored by DARPA.
The work of Yujia Zheng and Kun Zhang is supported by NSF Grant 2229881, the National Institutes of Health (NIH) under Contract R01HL159805, and grants from Apple Inc., KDDI Research Inc., Quris AI, and Florin Court Capital.


\bibliographystyle{plain}
\bibliography{reference}

\appendix

\newpage

\section{Sample Input Files}
\label{appendix:input}
\subsection{Trace File}
Following is a sample trace file from ModelServingS dataset. This can be automatically extracted from traces of requests and system deployment configs.
\begin{verbatim}
{
    "request.Client": {
        "service_description": "client that sends request to the router",
        "resources": {},
        "callees": [
            "Client-Router"
        ]
    },
    "request.Client.Client-Router": {
        "service_description": "network communication that
            send request from client to router",
        "resources": {},
        "callees": [
            "Router"
        ]
    },
    "request.Client.Router": {
        "service_description": "router that processes and 
            dispatches request to different servers",
        "resources": {},
        "callees": [
            "Router-Queue_0"
        ]
    },
    "request.Client.Router-Queue_0": {
        "service_description": "network communication that 
            send request from router to servers",
        "resources": {},
        "callees": [
            "Queue_0"
        ]
    },
    "request.Client.Queue_0": {
        "service_description": "when requests are received at a server, 
            they are buffered in the queue and waited to be executed. 
            Requests will be dequeued and processed by the batcher 
            when resources are available",
        "resources": {},
        "callees": []
    },
    "request.Client.Batcher_0": {
        "service_description": "when resources are available, 
            the batcher will check the queue and create a batch of 
            min(available requests, max batch size) requests 
            and send it to the model inference service",
        "resources": {},
        "callees": [
            "Queue_0"
        ]
    },
    "request.Client.ModelInference_0": {
        "service_description": "model inference service that 
            runs the model inference on the batched requests",
        "resources": {
            "GPU_0": "A A100 GPU"
        },
        "callees": [
            "Batcher_0",
            "ModelInference_0-Client"
        ]
    },
    "request.Client.ModelInference_0-Client": {
        "resources": {},
        "service_description": "network communication that 
            send the model inference result back to the client",
        "callees": []
    }
}
\end{verbatim}
\subsection{Measurement Description}
Following is a measurement description file explaining all the available metrics for different kinds of system components. These information can usually be found from documentation of observability framework and the observed system.
\begin{verbatim}
{
    "Client": {
        "request_level": {
            "latency": "The time taken for a request to be processed from 
                the time it is sent to the time the response is received"
        }
    },
    "Client-Router": {
        "request_level": {
            "latency": "The time taken for an invoked service to 
                process the request"
        }
    },
    "Router": {
        "request_level": {
            "latency": "The time taken for an invoked service to 
                process the request"
        },
        "service_level": {
            "throughput": "The number of requests that 
                a service successfully processes per second"
        }
    },
    "Router-Queue": {
        "request_level": {
            "latency": "The time taken for an invoked service to 
                process the request"
        }
    },
    "Queue": {
        "request_level": {
            "latency": "The time waited by a request in the queue 
                before it is dequeued",
            "queue_length": "The number of requests waiting in the queue 
                when a request is enqueued"
        },
        "service_level": {
            "enqueueing_rate": "The rate at which requests  are being 
                enqueued, in requests per second",
            "dequeueing_rate": "The rate at which requests are being 
                dequeued, in requests per second"
        },
        "to_exclude": {
            "service_level": [
                "throughput"
            ]
        }
    },
    "Batcher": {
        "service_level": {
            "max_batch_size": "The maximum size of 
                a batch that can be created"
        }
    },
    "ModelInference": {
        "request_level": {
            "latency": "The time taken for 
                an invoked service to process the request"
        },
        "service_level": {
            "execution_batch_size": "The number of 
                requests that are processed in a single batch",
            "throughput": "The number of requests that 
                a service successfully processes per second"
        }
    },
    "ModelInference-Client": {
        "request_level": {
            "latency": "The time waited by a request in 
                the queue before it is dequeued"
        }
    },
    "GPU": {
        "resource_level": {
            "utilization": "The percentage of time that 
                a resource is busy processing requests"
        }
    }
}
\end{verbatim}
Note that \texttt{to\_exclude} is used to help \sys{} generate a more accurate causal graph. In our context, throughput is not a relevant metric for the queue service since we already have the enqueueing rate and dequeueing rate as metrics.
\subsection{Common Metrics}
This file is used to realize metric enumeration for a more complete graph construction as discussion in \S\ref{sec:observation}.
\begin{verbatim}
{
    "service": {
        "request_level": {
            "latency": "The time taken for an invoked service 
                to process the request"
        },
        "service_level": {
            "throughput": "The number of requests that 
                a service successfully processes per second"
        }
    },
    "resource": {
        "resource_level": {
            "power": "A measure of capacity of 
                a resource to process requests",
            "utilization": "The percentage of time that 
                a resource is busy processing requests"
        }
    }
}
\end{verbatim}

\section{Sample Prompt}
\label{appendix:prompt}
Here is an sample prompt for the "Queue\_0" to figure out the causal relationship between its "queue\_length" and "dequeueing\_rate".

\begin{quotation}
[

\{'role': 'system', 'content': 'You are a service named \textbf{Queue} in a software system and your job is: \textbf{when requests are received at a server, they are buffered in the queue and waited to be executed. Requests will be dequeued and processed by the batcher when resources are available.} You are about to figure out the causal relationship between a metric of you and a metric of another system component.'\}, 

\{
'role': 'user', 'content':
\textbf{queue\_length} is a metric of you and it means the \textbf{number of requests waiting in the queue when a request is enqueued}. \textbf{dequeueing\_rate} is another metric of you and it means \textbf{the rate at which requests are being dequeued, in requests per second.}
If we can only choose one, which of the following cause-and-effect relationship is more likely?

A. A change in "queue\_length" of you directly causes a change in "dequeueing\_rate" of Queue\_0.

B. These two metrics do not directly influence each other, even if they might be correlated through other components in the system.

C. A change in "dequeueing\_rate" of Queue\_0 directly causes a change in queue\_length of you.

Please think step by step to make sure that you have the right answer.
Please select from one of the following options: [A, B, C] as the final answer. 
Put it as the only content in the last line.\}

]
\end{quotation}

Here is another sample prompt for the "ModelInference" to figure out how its "execution\_batch\_size" influence the max\_batch\_size of "Batcher". 
\begin{quotation}
[
{'role': 'system', 'content': 'You are a service named \textbf{ModelInference} in a software system and your job is: \textbf{model inference service that runs the model inference on the batched requests}.You are about to figure out the causal relationship between a metric of you and a metric of another system component.'}, 
{'role': 'user', 'content': '\textbf{Batcher\_0} is the next service that requests will invoke after you finish processing them, whose job is: \textbf{when resources are available, the batcher will check the queue and create a batch of min(available requests, max batch size) requests and send it to the model inference service} \textbf{execution\_batch\_size} is a metric of you and it means \textbf{The number of requests that are processed in a single batch}. \textbf{max\_batch\_size} is a metric of the next service and it means \textbf{The maximum size of a batch that can be created}.
If we can only choose one, which of the following cause-and-effect relationship is more likely?

A. These two metrics do not directly influence each other, even if they might be correlated through other components in the system.

B. A change in "max\_batch\_size" of Batcher\_0 directly causes a change in execution\_batch\_size of you.

C. A change in "execution\_batch\_size" of you directly causes a change in "max\_batch\_size" of Batcher\_0.

Please think step by step to make sure that you have the right answer.
Please select from one of the following options: [A, B, C], as the final answer. Put it as the only content in the last line.'}
]

\end{quotation}

Note that each query are repeated multiple times with varied order of options until a majority vote agreement is reached.

\end{document}